\RequirePackage{fix-cm}
\documentclass[smallextended]{svjour3}       
\smartqed  

\usepackage{hyperref}
\usepackage{graphicx}
\usepackage{amsmath}
\usepackage{multirow}
\usepackage{hyperref}
\usepackage{float}

\usepackage{algorithm}
\usepackage{algpseudocode}
\usepackage{amsmath} 
\algrenewcommand{\algorithmiccomment}[1]{\hfill\textbf{//}\,#1}

\usepackage{subcaption} 

\usepackage{tikz}
\newcommand{\roundbox}[1]{
	\begin{center}
		\begin{tikzpicture}
		\node[draw=black, rectangle, rounded corners](box){
			\begin{minipage}{0.95\columnwidth}
			#1
			\end{minipage}
		};
		\end{tikzpicture}
	\end{center}
}

\begin{document}

\title{IRJIT: A Simple, Online, Information Retrieval Approach for Just-In-Time Software Defect Prediction
}


\author{Hareem Sahar \and Abdul Ali Bangash \and Abram Hindle \and Denilson Barbosa }

\institute{Hareem Sahar \at
	Department of Computing Science \\
	University of Alberta, Edmonton, AB, Canada\\
	\email{hareeme@ualberta.ca}
	\and
	Abdul Ali Bangash \at
	 Department of Computing Science\\
	 Queen's University, Kingston, ON, Canada\\
	\email{abdulali.b@queensu.ca}
	\and
	Abram Hindle \at
	Department of Computing Science \\
	University of Alberta, Edmonton, AB, Canada\\
	\email{abram.hindle@ualberta.ca}
	\and
	Denilson Barbosa\at
	Department of Computing Science \\
	University of Alberta, Edmonton, AB, Canada\\
	\email{denilson@ualberta.ca}
}

\date{Received: date / Accepted: date}

\maketitle

\begin{abstract}
 Just-in-Time software defect prediction (JIT-SDP) prevents the introduction of defects into the software by identifying them at commit check-in time. Current software defect prediction approaches rely on manually crafted features such as change metrics and involve expensive to train machine learning or deep learning models. 
 These models typically involve extensive training processes that may require significant computational resources and time. These characteristics can pose challenges when attempting to update the models in real-time as new examples become available, potentially impacting their suitability for fast online defect prediction. Furthermore, the reliance on a complex underlying model makes these approaches often less \textit{explainable}, which means the developers cannot understand the reasons behind models' predictions. An approach that is not \textit{explainable} might not be adopted in real-life development environments because of developers' lack of trust in its results. 
 To address these limitations, we propose an approach called IRJIT that employs information retrieval on source code and labels new commits as buggy or clean based on their similarity to past buggy or clean commits. IRJIT approach is \textit{online} and \textit{explainable} as it can learn from new data without expensive retraining, and developers can see the documents that support a prediction, providing additional context. By evaluating $10$ open-source datasets in a within project setting, we show that our approach is up to $112$ times faster than the state-of-the-art ML and DL approaches, offers explainability at the commit and line level, and has comparable performance to the state-of-the-art.

\end{abstract}

\section{Introduction}
\label{intro}
The later a defect is identified in the software development life cycle, the higher its impact and maintenance cost~\cite{planning2002economic}. To identify and prevent defects early on, Just-in-time (JIT) software defect prediction (SDP)~\cite{mockus2000predicting,kamei2012large} is proposed to identify risky commits at check-in time. Identifying risky commits at commit-time allows practitioners to dedicate their limited testing resources to review and fix the riskiest commits while the context and changes are still fresh in the authors' minds.

JIT-SDP models are commonly built using machine learning (ML) and deep learning (DL) techniques. Once trained, these models rely on a static understanding of the data distribution they were trained on. However, as we show later, over time, these models can become outdated, resulting in a decline in their performance. Consequently, retraining becomes necessary. Nevertheless, retraining models from the ground up can be a resource-intensive and time-consuming process~\cite{wu2020deltagrad,cabral2023investigation}. In practice, these overheads can render such models impractical for JIT-SDP, which is inherently an online learning problem~\cite{cabral2019class,tabassum2020investigation}.

ML and DL models are not only cost-prohibitive due to the time and resources needed for training but often lack explainability. \textit{Explainability} is a desired property of a prediction model designed for deployment in a real life development environment. An explainable approach must provide insights into the reasoning behind its decisions. For a JIT approach to be called \textit{explainable}, the developers shall be able to trace back predictions to their originating documents~\cite{dam2018explainable}. An example would be highlighting parts of a change that resemble a past buggy change. The state-of-the-art JIT-SDP approaches such as DeepJIT~\cite{hoang2019deepjit} and CC2Vec~\cite{hoang2020cc2vec} are not explainable because of reliance on a complex underlying learning model such as a deep neural network model. 
 
To address these limitations of prior work, in this paper, we propose IRJIT, a simple, online, and explainable information retrieval (IR) approach for JIT defect identification. Unlike existing methods~\cite{cabral2019class,pornprasit2021jitline} that may offer these features independently, IRJIT integrates these advantages into a single model. Additionally, IRJIT employs a unique classification strategy and classifies a new commit (as clean or buggy) based on its similarity to past commits, computed using the BM25 algorithm~\cite{BM25}.

For each identified buggy commit, IRJIT also estimates its buggy lines using \textit{term frequency × inverse document frequency} (tf–idf) term weights~\cite{salton1978term}. By relying on traditional document indexing using source code token frequencies instead of hand-crafted features such as code change or review metrics~\cite{mcintosh2017fix}, we make our approach readily deployable for any project with an active version control system and an issue tracker. Our approach offers several advantages over recent JIT approaches~\cite{hoang2019deepjit}~\cite{hoang2020cc2vec}:
\begin{itemize}
	\item IRJIT operates in an \textbf{online} manner, which means that it can adapt over time by learning from new changes. IRJIT incrementally updates indexes as new commits arrive without undergoing frequent and extensive retraining, which can slow down the overall prediction process. Being able to update online and react promptly to changes is desirable for the adoption of a JIT approach~\cite{wan2018perceptions}. By reducing the costs associated with retraining models, IRJIT quickly reacts to changes, making it ideal for JIT-SDP.
	
	\item IRJIT is \textbf{explainable} as it links predictions to their originating documents. It also ranks source code lines of identified buggy commits by riskiness, making it easier for developers to inspect source code documents that were used in prediction and understand why a commit was classified as buggy. These characteristics enhance transparency and increase practitioners' trust in  predictions~\cite{dam2018explainable}, while potentially simplifying the bug-fixing process.

	\item IRJIT is \textbf{simple} as it operates directly on source code without requiring additional hand-crafted features. It does not rely on complex machine learning or deep learning prediction models. Instead, it draws conclusions by matching current changes against the past source code documents that developers are already familiar with; hence, the results are also simpler to understand. 
  
\end{itemize}

To establish the strengths and weaknesses of our approach, we conduct a comparison with four state-of-the-art methods, including Oversampling Rate Boosting (ORB)~\cite{cabral2019class},  JITLine~\cite{pornprasit2021jitline}, JITFine~\cite{jitfine} and N-gram approach of Yan et al~\cite{yan2020JIT}. ORB is a leading online JIT-SDP approach that produces balanced recalls for both defective and clean classes by automatically adjusting the resampling rate to mitigate class imbalance. Conversely, JITLine engages in extensive tuning through SMOTE~\cite{chawla2002smote}
and Differential Evolution (DE) to directly optimize metrics and word vectors for AUC. This process could delay the \textit{online} predictions~\cite{tan2015online}\cite{cabral2019class} making the approach less suitable for the continuous integration pipelines. Furthermore, as a consequence of JITLine's reliance on SMOTE to create new training examples, it cannot retrieve originating commits to be blamed for a bug. In other words, it cannot explain the predictions by linking to the original documents causing the classification, although it can rank lines by defectiveness. Despite JITLine's high performance, it was outperformed by JITFine~\cite{jitfine} which uniquely combined defect prediction and localization into a unified deep learning (DL) model leveraging both semantic and expert features of the source code.

\section{Research Question and Contributions}
This paper investigates when and to what extent information retrieval can be helpful in a realistic online JIT-SDP scenario. We answer following Research Questions (RQs) in this paper:

\textbf{RQ1. Is the CPU run time of IRJIT less than (or equal to) that of a state-of-the-art machine learning approach?}\\
\textbf{Motivation.} For this RQ, cost is measured as CPU run time of building a model. The cost of using IR methods in JIT-SDP is not thoroughly studied. Additionally, the existing literature does not provide a clear answer regarding the trade-offs between simple versus heavily-trained SDP models. Answering RQ1 builds a clearer understanding of cost differences and allows organizations to allocate their resources more wisely, especially when budgets are limited. The insights could also drive the SDP research towards more cost-effective methods.

\textbf{RQ2. Does the predictive performance of IRJIT exceed (or remain equal to) the performance of a state-of-the-art machine learning approach?}\\
\textbf{Motivation.} This work introduces a JIT-SDP approach that adapts without retraining. The question arises: How does the simplicity of our approach translate to its performance? Does a straightforward approach hold its ground when compared against a sophisticated model? To answer this, we analyze the predictive performance of our approach and compare it with a machine learning baseline to see if it improves, deteriorates or retains similar predictive performance as the baseline approach. The insights from RQ2 will help stakeholders in choosing between our IR approach and the ML based state-of-the-art approach.

\textbf{RQ3. For fine grained line level SDP, does IRJIT performance exceed (or remain equal to) that of a state-of-the-art machine learning approach?}\\
\textbf{Motivation.} Fine-grained SDP helps developers identify bugs quickly with less effort. If our approach can predict buggy lines upon commit, it might expedite the bug-fixing process, enhancing the overall software development life cycle. On the other hand, if buggy lines are not ranked on top, developer effort may go to waste, and the bugs may still go unnoticed. Therefore, in this research question, we measure the performance of our approach in terms of correctly ranking buggy lines. The contributions of this work include:

\begin{itemize}
	\item{a JIT-SDP approach based on IR to detect buggy commits at commit check-in time;}
	
	\item{an approach to identify associated buggy lines for each identified buggy commit;}
	
	\item{a realistic evaluation of our approach on $10$ open-source software defect prediction datasets using various metrics and its comparison with a mature machine learning baseline.}
\end{itemize} 

Our replication package is available at \url{https://github.com/Hareem-E-Sahar/eseval\_online}.

\section{Background and Related Work}

\subsection{JIT-SDP}
\label{sec:1}
{JIT-SDP} emphasizes the identification of risky changes just-in-time, i.e., before they reach repository. Existing JIT approaches detect bugs based on past information such as change histories~\cite{buggyorclean} or by relating past defects to software metrics~\cite{commitguru}\cite{kamei2012large}.
CLEVER~\cite{clever} identifies new defects by comparing new code to known cases of buggy code. It uses a clone detection to improve the  accuracy of buggy change identification which was originally based solely on a metric-based model. Kamei \emph{et al.}~\cite{kamei2012large} performed a longitudinal case study on two open-source systems, Openstack and Qt, and found that fluctuations in the properties of fix-inducing changes can impact the performance of JIT models. They also showed that JIT models typically lose power after one year, so they should be updated with more recent data. 

\begin{figure*}
	\centering
	\includegraphics[width=0.9\linewidth]{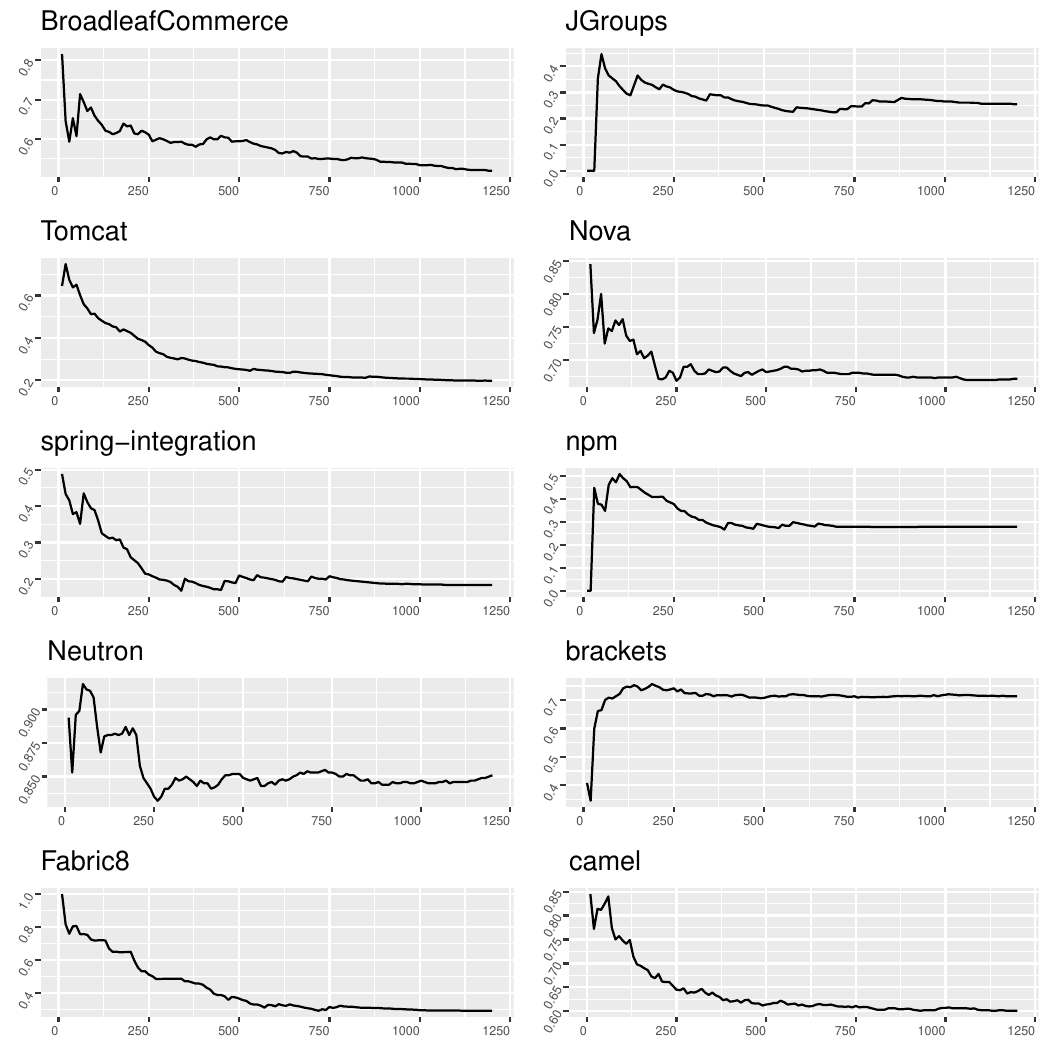}
	\caption{Performance evolution of a typical JIT-SDP approach over time showing a declining trend. x-axis shows time step and y-axis shows G-mean values. The model was trained once and the subsequent predictions were made without retraining the model.}
	\label{fig:prerqplot}
\end{figure*}

Figure~\ref{fig:prerqplot} illustrates an example of one such machine learning based SDP model, trained on the initial 10\% project data. 
The plot reveals a notable decline in the model's performance over time, a phenomenon known as concept drift in the literature~\cite{kamei2012large,cabral2019class}, which underscores the importance of retraining models to ensure prediction accuracy.
This need for regular updates brings into question the costs of retraining. If these costs are high, they could hinder the model's practical application, emphasizing the necessity for computationally efficient models. In this context, the contributions of Pornprasit \emph{et al.} are particularly significant. Notably, Pornprasit \emph{et al.}~\cite{pornprasit2021jitline} demonstrated that their machine learning model outperforms complex deep learning methods like JITFine~\cite{jitfine} DeepJIT~\cite{hoang2019deepjit} and CC2Vec~\cite{hoang2020cc2vec}, which require significant computational resources. 
Our proposed method also offers reduced computational costs by eliminating the need for full retraining, making it a practical solution for JIT-SDP.

\subsection{Online JIT-SDP}
Online learning has been extensively studied and applied in recent years~\cite{cabral2023investigation,tabassum2020investigation}. It refers to learning from data streams where data arrives continuously and a timely response is required. Strictly speaking, online learning algorithms process each training example once ``on arrival" without the need for reprocessing, and maintain a model that reflects the current state to make a prediction at each time step. Tan \emph{et al.}~\cite{tan2015online} investigated JIT-SDP using batches of training examples arriving over time to update the prediction models. They used resampling techniques to deal with the class-imbalanced data issue and updatable classification to learn over time. They also considered the fact that the labels of training data may arrive much later than the commit time, a concept named verification latency in literature~\cite{cabral2019class}. Cabral \emph{et al.}~\cite{cabral2019class} were the first to propose an online approach for JIT-SDP. Their approach called Oversampling Rate Boosting (ORB) tackles concept drift (change in the proportion of defect-inducing and clean examples) in an online JIT-SDP scenario, while taking verification latency into account. ORB automatically adjusts the re-sampling rate to tackle class imbalance, which improved its predictive performance over JIT approaches that assume a fixed level of class imbalance. Tabassum \emph{et al.}~\cite{tabassum2020investigation} evaluated ORB in a cross-project (CP) defect prediction setting and showed that CP models trained together on CP and within project (WP) data have a higher performance than WP models trained only on WP data.

\subsection{Explainability}
Explainability in SDP aims at making the outputs of defect prediction models more understandable to humans. This aspect is crucial to JIT-SDP models because developers favor simple, explainable, and online models, as highlighted in the survey conducted by Wan \emph{et al.}~\cite{wan2018perceptions}. Traditional defect prediction models leverage ML algorithms to identify potential defects based on historical data. However, the complexity and ``black box" nature of these models often make it difficult for practitioners to understand and trust their predictions. Pornprasit \emph {et al.}~\cite{pornprasit2021jitline} addressed this critical shortcoming of ML and DL models i.e., their lack of explainability, by utilizing a model agnostic technique, LIME~\cite{ribeiro2016should}, to explain the reasoning behind the decisions of their commit level random forest classifier. Wattanakriengkrai \emph{et al.}~\cite{wattanakriengkrai} applied the same technique to explain a file level defect model. The recognition of the need for explainable models is growing among researchers. In line with this, our goal is to develop an approach that enhances explainability by highlighting the historical instances that inform its predictions.

\subsection{Verification latency in JIT-SDP} Veriﬁcation latency refers to the fact that the labels of training examples may arrive much later than their input features. Ignoring such delay in an online learning scenario means training models on data that is not available in practice. In JIT-SDP a commit could receive its true label early if a defect associated to it is quickly found. Alternatively, it can receive label at the end of the waiting time (if the commit is believed to be clean) or after the waiting time (if an example previously considered clean is found to actually be defect-inducing after the waiting time has passed). This waiting time should reﬂect the time it takes for one to be conﬁdent enough that software changes are not defect-inducing. Tan \emph{et al.}~\cite{tan2015online} introduced this concept and found that ignoring latency could lead to false performance estimates. Cabral \emph{et al.}~\cite{cabral2019class} investigated how long it typically takes for software changes to be revealed as defect-inducing. Later, Tabassum \emph{et al.}~\cite{tabassum2020investigation} and our work followed the same methodology for the evaluation of JIT-SDP.

\subsection{Leveraging source code similarity}
\label{sec:codesim}
DeepBugs\cite{deepbugs} detects incorrect code by reasoning about identifier names, and training a binary classifier on correct and incorrect examples. We leverage code similarity between past buggy and clean changes to \textit{explain} why a new change is classified as buggy. This line of work is inspired by the success of code search and recommendation engines~\cite{aroma}\cite{facoy} which employ IR techniques. IR-based methods work by matching shared vocabulary between query and corpus documents, and are effective for code retrieval~\cite{sourcerer2009}\cite{lancer} and bug localization~\cite{rahman2018improving}. Due to the repetitive nature of source code~\cite{hindle2012}, statistical language models can also spot defective code effectively~\cite{ray2016,campbell2014syntax,santos2018syntax}. Yan \emph{et al.}~\cite{yan2020JIT} use several change level features and software naturalness with the N-gram model to localize buggy changes. Their proposed framework ranks buggy lines in order of suspiciousness. In this work, we also estimate the riskiness of the source code lines based on the number of occurrences of buggy tokens in each line. 

\section{IRJIT in Practice}
Here we are going to share the story of how IRJIT fits into Ada's workflow who is a software developer at TechFlow\footnote{It is a hypothetical name}. Every day, Ada faces a challenge to ensure that her code commits are defect-free. While comprehensive testing processes and code reviews are a part of TechFlow's quality control, the idea is to catch potential issues as early as possible, minimizing disruptions and maximizing efficiency. IRJIT is seamlessly integrated into the developers' workflow using a commit hook, preventing potential defects from sneaking into the codebase, hence making Ada's life easier. Imagine Ada had just finished writing a feature that involved intricate logic. As she prepared to commit her changes, the newly-integrated JIT prediction model sprang into action. Behind the scenes, the IRJIT model analyzes various aspects of her commit such as the source code changes made to different files in commit. These changes, along with the historical commit data continuously collected by IRJIT, serve as the foundation for the model's prediction. By analyzing previous commits and their labels, the model learns characteristics that make a commit more or less likely to introduce defects. A moment later, a notification popped up, suggesting that there might be a potential issue with Ada's commit. Intrigued, Ada decided to investigate further by looking at the past changes that matched the new changes, and the ranked buggy lines. She realized that she had missed handling an edge case in her code. Grateful for the timely alert, she made the necessary modifications before proceeding with the commit.  

For this scenario to work, IRJIT must be online and update itself in real-time as new commits and defect labels arrive. To accomplish this, IRJIT operates in the background, automatically retrieving data from various sources, including version control and issue-tracking systems. The new commits are placed in a queue and wait for a defect to be linked to them or for a set time period to pass, after which we assume the commit is clean. As Ada's team fixes bugs that arise post-commit, IRJIT updates its understanding of which commits led to defects. Once IRJIT obtains a new labeled commit, its prediction model is updated incrementally without requiring complete retraining from scratch. This online approach taken by IRJIT ensures that developers' workflow remains uninterrupted.

As days turned into weeks, Ada, along with her colleagues at TechFlow, found the IRJIT becoming an indispensable part of their workflow. There were moments of affirmation where the model caught potential issues. While the model was not infallible, it allowed Ada and her team to commit code with an added layer of confidence. There were also moments when Ada disagreed with the model's prediction, and in such cases, Ada could ignore the prediction and still push the commit. This balance between automated insights and human judgment ensured that the tool remained an assistant, not a gatekeeper.

\section{IRJIT Methodology}
\label{sec:methodology}
\begin{figure}
	\centering
	\includegraphics[scale=0.99]{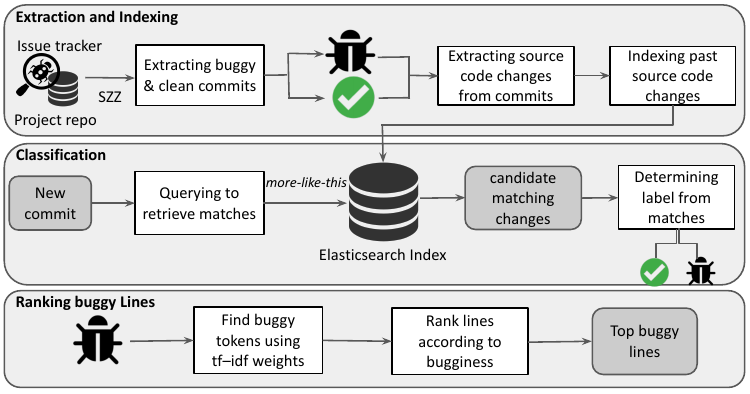}
	\caption{An overview of IRJIT operation. IRJIT extracts past changes using SZZ and indexes them using inverted indexes. On arrival of a new commit, IRJIT classifies the commit based on its similarity with the past changes in the index. Finally, IRJIT ranks changed lines according to bugginess.}
	\label{fig:methodology}
\end{figure}

IRJIT can be integrated into git either through a commit hook~\cite{hooks} or a pull request bot. As soon as a developer pushes a new commit, IRJIT analyzes the new source code changes and compares them against past changes in the project's version control system. The developer is alerted if IRJIT predicts the new commit as buggy. This way, IRJIT can ensure that bugs are identified and resolved before the changes reach the main branch of the project repository. Figure~\ref{fig:methodology} shows the backend design of IRJIT framework. IRJIT extracts past project commits (Section~\ref{sec:szz}) and associated source code changes (Section~\ref{sec:diff}). The source code changes are stored into an inverted index for querying (Section~\ref{sec:index}). IRJIT intercepts new commits, queries an index for matches in the indexed corpora, and flags changes that match known buggy changes (Section~\ref{sec:classify}). Lastly, for each buggy change, IRJIT ranks source code lines that are part of the change according to their riskiness (Section~\ref{sec:ranklines}).

\subsection{Extracting buggy and clean commits}
\label{sec:szz}
As a first step, IRJIT requires a corpus of buggy and clean commits to make predictions. To obtain buggy commits from a project's history IRJIT relies on SZZ algorithm~\cite{szz2005,szz2006}, which operates on the data available in the version control and issue tracking system. In the first step, the algorithm identifies a defect-fixing commit. A commit is \textit{fixing} if it fixes a defect, i.e., it refers to an issue labeled as \textsc{fixed} in the issue tracking system. In the second step, SZZ traces back through the version control to identify the associated \textit{buggy} commit i.e. the one responsible for inducing the defect that the fixing commit addressed.
For this purpose, SZZ algorithm leverages \textit{diff} and \textit{annotate/blame} functionality. The \textit{diff} helps identify lines that were changed between a defect-fixing commit and its immediately previous commits. These are the lines that potentially fixed the bug. The \textit{annotate/blame} functionality identifies commits that originally introduced or modified those lines in the past. SZZ flags those original changes as defect-introducing or buggy if the commit was made before filing the issue in project issue tracker.
Once the algorithm completes its execution, the buggy commits are identified. Similar to prior work~\cite{yan2020JIT}\cite{ISSTA2021}, we include the remaining commits in IRJIT's analysis as clean commits. From here onwards, we only use the terms \textit{buggy} and \textit{clean} to refer to commits.

\subsection{Extracting source code changes from commits}
\label{sec:diff}

A commit may introduce source code changes across multiple files. A \textit{change} refers to committed code in a single file. IRJIT extracts changes using the \textit{git diff} utility and pre-processes them to remove blank lines.

The processed source code changes for each file are subsequently stored in a JSON document representing that file. To facilitate the retrieval of specific modifications later, we opt not to consolidate changes from different files, maintaining the separation of changes specific to individual files instead. Within each JSON document, essential metadata such as \texttt{lines added}, commit hash, file name, and label are included. 
Currently, IRJIT collects and uses \texttt{lines added} by default because SZZ algorithm only flags changes that introduce new lines as buggy~\cite{mcintosh2017fix}. Purushothaman \emph{et al.}~\cite{test} also suggests that the probability of introducing a bug is higher with the addition of new lines. Hence, only \texttt{lines added} field might be enough to spot bugs. Our preliminary experiments~\footnote{\url{https://github.com/Hareem-E-Sahar/eseval_online/tree/main/eseval_timewise_batched/results_commit_level/results_lines_added_deleted_camel}} also showed that analyzing \texttt{lines deleted} did not lead to a significant gain in the predictive performance of our model. Therefore, we only use the \texttt{lines added} field to build the IRJIT model.

\subsection{Indexing past source code changes}
\label{sec:index}
To make the classification of a new commit scalable and efficient, IRJIT employs an inverted index, a data structure designed to maintain a terms to documents mapping. In information retrieval tasks, inverted index helps to optimize search and retrieval from a large corpus of documents. To build inverted indexes, IRJIT leverages Elasticsearch, which is a real-time full-text search engine built on top of the Apache Lucene library~\cite{lucene}. 

The historical source code changes are tokenized and indexed based on user specified settings. As mentioned Section~\ref{sec:codesim}, prior works employ source code as features for various tasks. IRJIT uses a whitespace tokenizer~\cite{whitespace} which is part of our custom camel-case analyzer~\cite{pattern} to tokenize the past source code changes, specifically the \texttt{lines added} to a change. It splits the words into tokens when a case change is encountered using a method provided in the ElasticSearch documentation. The choice of a camel-case analyzer was initially influenced by Campbell's work~\cite{campbell2016unreasonable}. However, the final selection was based on the finding that it performs slightly better than other analyzers, such as simple analyzer and the shingle analyzer~\footnote{\url{https://github.com/Hareem-E-Sahar/eseval_online/tree/main/eseval_timewise_batched/results_commit_level/results_lines_added_shingle}} 

In practice, one index per project can serve the purpose of a repository for all prior changes. The index is initially empty, resembling a predictive model that will always predict clean until a buggy commit arrives. It is then updated incrementally by adding changes when they arrive, eliminating the need of re-indexing any previous documents. However, when a software change arrives, it is saved in a queue for an arbitrary amount of time, referred to as the waiting period. This waiting period serves to establish confidence that the change is clean. If a bug is associated with the change during the waiting period, it immediately becomes a buggy example. Otherwise, the change becomes a clean example at the end of the waiting period.
IRJIT's approach empowers developers to query against the new changes right away, unlike existing approaches that necessitate model retraining when new data arrives. The ability to remain online makes IRJIT suitable for repositories of any size.

\subsection{Classifying a new commit}
\label{sec:classify}
\subsubsection{Retrieving similar commits}
\label{sec:MLT}
When a new commit arrives to be checked-in to a version control system, IRJIT treats it as a test commit. The camel-case analyzer is used to process the commit, similar to the indexed commits, before classifying it as either buggy or clean. This classification is based on the extent to which the new commit's changes match with those from past commits.
The classification process involves several steps. First, IRJIT extracts source code changes from the new commit. Second, for each extracted change it formulates a query using the \texttt{lines added} field from the corresponding change. Third, it executes the query and retrieves matches from the past source code changes. Specifically, IRJIT formulates a \textit{more\_like\_this} (MLT) query~\cite{MLT}, which finds indexed training documents that are ``like" a given test document. MLT query\footnote{min\_term\_freq=1, and   min\_doc\_freq=1} utilizes corpus-level and document-level term statistics, i.e., tf–idf to compute similarity between a query and documents in the corpus. \textit{tf} represents the frequency that a source code token appears in a specific document, and \textit{idf} represents the universal importance of a source code token for the entire collection of documents. IRJIT utilizes BM25 algorithm\cite{BM25} to calculate similarity. BM25 is a scoring function that considers term frequencies, document frequencies, and document length to generate a relevance score. 

Notably, the query document and the indexed documents are processed using the same analyzers before computing similarity. Although Elasticsearch allows for customization of similarity algorithms, we found the default BM25 algorithm to be adequate for our study. Based on the BM25 similarity results, Elasticsearch returns the top-K most similar training documents for a given test document along with their respective relevance scores.

IRJIT may issue multiple queries for each test commit, depending on the number of changes within that commit. The matching documents from all MLT queries are combined into a set called candidate matches and sorted according to relevance score. The Elasticsearch scoring formula~\cite{scoring} is originally based on tf–idf and BM25, but contains adjustments to allow for comparison of scores across different queries~\cite{campbell2016unreasonable}. Next, IRJIT's classifier determines a final label for the test commit.

\subsubsection{Determining the label from matches} 
\label{sec:knn}
IRJIT currently implements a K-nearest neighbor (\texttt{KNN}) classifier which classifies a commit as \textit{buggy} or a \textit{clean} based on the majority label of the candidate matches. If the majority of K neighbors are buggy, the test commit is labeled buggy; otherwise, it is considered clean. The choice of K is crucial and can be fine-tuned for each project to optimize performance. Alternatively, the user can provide this value as input. Currently, IRJIT uses a default value of $K=3$ which was set empirically after experimenting on $K={1,3,5,7,9,11}$.

\subsection{Generating a ranking of buggy lines for a commit}
\label{sec:ranklines}
If a new change is identified as buggy, IRJIT provides developers with a ranking of modified lines based on their level of bugginess. Prior work~\cite{hindle2012}\cite{ray2016}\cite{yan2020JIT} exploits N-gram models to find most entropic source code tokens. However, we adopt a slightly different approach to estimate the buggy tokens, i.e., tokens that contribute towards overall bugginess of the document. We rely on the BM25 term weights obtained while executing \textit{more\_like\_this} query (Section~\ref{sec:MLT}). These term weights highlight the significance of individual terms when predicting a document as buggy. The terms with highest weight when a document is predicted buggy are considered buggy tokens. These tokens are indicative of potential sources of bugs. After extracting buggy tokens, we rank lines based on the number of occurrences of buggy tokens within each line. Finally, we show the ranked buggy lines to the developer. Our ranking approach allows for a unified model for JIT prediction and localization, eliminating the need for separate models for each task. This integration ensures that explanations are generated promptly alongside predictions, essential for real-time application in development workflows.

\section{Experimental Setup}
\subsection{Dataset}
We used a dataset extracted by Cabral \emph{et al.}~\cite{cabral2019class}, which contains ten existing open-source projects.
The dataset is available at \url{https://zenodo.org/record/2594681}. The size of the datasets range from 74k to over 1.3m lines of code, and their development periods span from $6$ to $14$ years. The defect datasets contain commit-level metrics proposed by Kamei \emph{et al.}~\cite{kamei2012large} and used to describe a software change. The metrics can be divided into five categories including i) diffusion, ii) size, iii) purpose, iv) history, and v) developer experience. All the datasets were extracted using CommitGuru. Table~\ref{table:dataset} shows the number of commits, percentage of defective commits, development time period, and programming language for each project.

\begin{table*}
	\caption{Details of the studied projects}
	\label{table:dataset}
	\begin{tabular}{lrccc}
       \hline
       \noalign{\smallskip}

		Projects & \#Commits&\% Defective & Period & Language\\
		\noalign{\smallskip}\hline\noalign{\smallskip}

        \hline
  
		Brackets           & 17,310 & 23\% & 12/2011 - 12/2017   & Java script\\
		BroadleafCommerce  & 14,910 & 17\% & 09/2003 - 12/2017   & Java\\
		Camel              & 30,517 & 20\% & 03/2007 - 12/2017   & Java\\
		Fabric8            & 13,003 & 20\% & 12/2011 - 12/2017   & Java\\
		JGroups            & 18,316 & 17\% & 09/2003 - 12/2017   & Java\\
		Neutron            & 19,451 & 24\% & 12/2010 - 12/2017   & Python\\
		Nova               & 48,937 & 25\% & 08/2010 - 01/2018   & Python\\
		Npm                & 7,892  & 18\% & 09/2009 - 11/2017   & Java script\\
		Spring-integration & 8,691  & 27\% & 11/2007 - 01/2018   & Java\\
		Tomcat             & 18,877 & 28\% & 03/2006 - 12/2017   & Java\\
  \noalign{\smallskip}\hline

	\end{tabular}
\end{table*}

\subsection{Training and Prediction}

JIT-SDP is an online process that involves dealing with commits that arrive continuously over time. Each commit has a software change associated with it, which allows us to classify the commit as defect-inducing or clean. A time step associated with the commit indicates the order of arrival of commit based on the author's timestamp. The model's predictive performance is measured at a specific time step \textit{t} by training it on commits received before \textit{t} and then evaluating it on the commit that arrives at \textit{t}. During the initial phase of a project, the model predicts a test commit as clean until it encounters a defect-inducing example. Therefore, one must update the model as new commits and their labels become available.

We evaluate our approach in the context of within-project defect prediction~\cite{tabassum2020investigation} setting. We use past data from the same project to build our models, or in other words, each model is trained and tested on the same project. We ensure that our evaluation avoids time-travel~\cite{bangash2020time} by preserving the chronological order of data. This is achieved by making certain that our test data always follows the training data. For a JIT model, it's essential to adapt online~\cite{cabral2019class,tabassum2020investigation}, which means learning from new data as soon as it becomes available without the need to retrain on the old data. IRJIT has the ability to update in real time with new data.

\subsection {Baselines} 

We compare IRJIT with four state-of-the-art approaches from the literature, the online method known as Oversampling Rate Boosting (ORB)~\cite{cabral2019class}, two offline approaches --- JITLine~\cite{pornprasit2021jitline}, JITFine~\cite{jitfine}, and Yan et al's N-gram based defect localization model~\cite{yan2020JIT}. Cabral \emph{et al.} were the first to address JIT-SDP as an online learning problem considering verification latency. ORB~\cite{cabral2019class} automatically adjusts its resampling rate to mitigate the evolving class imbalance, thus ensuring balanced recall rates for both buggy and clean classes.

Pornprasit \emph{et al.} proposed JITLine~\cite{pornprasit2021jitline} which constructs a ML model on source code changes plus expert features. To counteract the data imbalance problem caused by a low percentage of buggy commits, JITLine generates new documents leveraging SMOTE technique that is optimized using the Differential Evolution algorithm. We selected JITLine as our baseline because of its superior performance, surpassing various deep-learning methods such as CC2Vec~\cite{hoang2020cc2vec}, and DeepJIT~\cite{hoang2019deepjit}, as well as Yan \emph{et al's.}~\cite{yan2020JIT} N-gram model. To the best of our knowledge there is no other ML approach that has outperformed JITLine. Additionally, JITLine provides fine-grained predictions by ranking buggy lines of an identified buggy commit. It uses LIME to build a local model around the predicted instance which then generates ranking of risky lines. In contrast, IRJIT uses the importance score of terms involved in the prediction of a commit to identify buggy lines.

Yan \emph{et al's}~\cite{yan2020JIT} notable contribution was a fine-grained model for ranking source code lines by their defect proneness. They used an existing JIT-SDP technique based on expert features to identify buggy commits and then apply their N-gram based defect localization model to rank lines within the identified buggy commits. We include a comparison of their N-gram model with IRJIT in this paper. Ni \emph{et al.} introduced JITFine~\cite{jitfine}, a unified model that integrates semantic and expert features of source code through a deep learning technique, CodeBERT, thus eliminating the need for a separate defect localization model. JITFine outperformed JIT-Fine and all other models on both defect prediction and localization tasks. Similar to JITFine, IRJIT uses a single model to achieve both defect prediction and localization.

\subsection{Evaluation Metrics}
\label{sec:metrics}
In JIT-SDP, which is an online learning problem, the most frequently used metrics are Recall$_{0}$, Recall$_{1}$ G-mean~\cite{cabral2019class}. 

\textbf{Recall$_{0}$} (R${_0}$) measures the proportion of clean class examples that are correctly classified as clean. It is also called true positive (TN) rate. It is calculated as:

\begin{displaymath}
Recall_{0} = \frac{TN}{TN+FN}
\end{displaymath},

\textbf{Recall$_{1}$} (R${_1}$) measures the proportion of buggy class examples that are correctly classified as buggy. It is also called true positive (TP) rate. It is calculated as:

\begin{displaymath}
Recall_{1} = \frac{TP}{TP+FN}
\end{displaymath},

\textbf{G-Mean} is the geometric mean of positive accuracy (TP rate) and negative accuracy (TN rate). A good classifier will exhibit high accuracies for both classes, resulting in a high G-mean. G-mean is calculated as shown below:
\begin{displaymath}
G-mean=\sqrt{\frac{TP}{TP+FN} * \frac{TN}{TN+FP}}
\end{displaymath}

For imbalanced learning problems, such as defect prediction, it's recommended to use metrics that are not sensitive to class imbalance, such as \textbf{G-mean} and the difference of recalls \textbf{$|$R0-R1$|$}. According to Wang and Minku~\cite{wang2018systematic} G-mean helps ensure fair evaluations. It is important to note that metrics sensitive to class imbalance, like Precision and F-measure, can lead to inconsistent performance evaluations~\cite{gracia2009learning} as class distribution varies across datasets. In other words, the performance measure will change as class distribution changes, even though the underlying performance of the classifier does not. Lastly, the AUC (Area Under the Curve) metric relies on varying the classification decision threshold for separating positive and negative classes in the testing dataset. In other words, calculating AUC requires a set of confusion matrices~\cite{wang2018systematic}. Therefore, unlike other measures based on a single confusion matrix, AUC cannot be used as an evaluation metric in online learning without memorizing data. For this, reason we do not use AUC in our evaluation.

For line-level evaluation, we employ the performance metrics used in the previous fine-grained software defect prediction studies~\cite{pornprasit2021jitline,jitfine,yan2020JIT}. \textbf{\textit{TopK}} accuracy measures the ratio of actual buggy lines to the Top \textit{K} lines. A high ratio suggests that many actual buggy lines are ranked at the top. \textbf{\textit{Recall@20\%Effort}} measures the ratio of actual buggy lines that can be found given only top 20\% changed lines of code. In other words, the top K\% of lines are ranked by their defect-proneness. \textbf{\textit{Effort@20\%Recall}} measure the amount of effort made by the developers to identify top20\% of buggy lines in a commit. A higher value means that more effort is needed. \textbf{\textit{Initial False Alarm (IFA)}} measures the number of clean lines in a commit that need to be inspected before identifying an actual buggy line.  This means developers need to spend effort on IFA number of lines until the first buggy line is found once lines are ranked by their bugginess. Ideally buggy lines should be within the first few inspected lines.

\subsection{Evaluation methodology}
\label{sec:eval}

IRJIT uses information, known by time step \textit{t} only, to predict a commit at \textit{t}. Correspondingly, our models update in real-time with the recent information available at each time step. The time step is just a sequential number signifying the order in which we receive project data. To predict a commit that arrives at \textit{t}, the prediction model incorporates all previously labeled training data available by the Unix timestamp corresponding to \textit{t}. For an online learning problem~\cite{tabassum2020investigation} such as JIT-SDP, this kind of evaluation is more realistic than an 80:20 split. It aims to replay an online setting where changes continuously arrive over time, and predictions are only based on past changes. Consequently, performance estimates align more closely with what developers may encounter in a real-world software development environment.
\begin{algorithm}
	\caption{IRJIT's Approach}\label{algo:irjit}
	\begin{algorithmic}[1]
		\State \textbf{Input:} $d$ = incoming training examples, $w$ = waiting period, $K$ = \# of neighbors to consider for the Knn algorithm, 
		\State Create index $idx$
		\State $\hat{y} = \text{`False'}$
		\For{each training example $d_t = (x_t, y_t)$, $t \gets 0$ to $\infty$}
		\State $matches = \text{MLT\_query}(x_t)$
		\State $\hat{y} = \text{KNN\_clf}(matches, K)$
		\State $\text{update\_WFLQ}(d_t)$ \Comment{incoming example stored in queue, waiting to be used for training}
		\For{each item $q_i$ in WFLQ}
		\If{a defect was linked to $q_i$ at a timestamp $\leq t$}
		\State create a buggy $training\_example$ for $q_i$
		\State $\text{index}(idx, training\_example)$
		\State remove $q_i$ from WFLQ
		\Else
		\If{$q_i$ is older than $w$}
		\State create a clean $training\_example$ for $q_i$
		\State $\text{index}(idx, training\_example)$
		\State remove $q_i$ from WFLQ
		\State $\text{update\_CLQ}(training\_example)$ \Comment{CLQ stores clean training examples}
		\EndIf
		\EndIf
		\EndFor
		\If{a defect was linked to a $training\_example$ in CLQ\\ at a timestamp $\leq t$}
		\State Swap the label of $training\_example$ to defect-inducing
		\State $\text{index}(idx, training\_example)$
		\State remove $training\_example$ from CLQ
		\EndIf
		\EndFor
	\end{algorithmic}
\end{algorithm}

We ran our experiments $10$ times on each project shown in Table~\ref{table:dataset}. We evaluated IRJIT following Cabral \emph{et al's}~\cite{cabral2019class} online methodology, which takes \textit{verification latency} into consideration. 
Algorithm~\ref{algo:irjit} presents our evaluation methodology. The algorithm accepts three inputs, d: Incoming training examples,K: number of nearest neighbors to consider in the KNN classification, and, w: the waiting period in the queue for the arrival of labels.

Initially an Elasticsearch index is created with no data in it. When a new change d$_t$ is received at time step t (line $4$), it is presumed to be non-defect-inducing. For this change d$_t$, IRJIT retrieves matches by executing a more like this query (line $5$) based on the source code lines added to a change, represented by x$_t$. Then, the KNN algorithm (line $6$) is applied to the matches to obtain a prediction for the change based on the majority consensus of the retrieved matches. The queue is also updated to store d$_t$. The change d$_t$ remains in the queue waiting for its label to be assigned. The labels are assigned to the stored examples when the queue is processed (line $8$ to $21$). During queue processing, the algorithm  iterates over each change q$_i$ and determines whether it has received its label.
If a defect is linked to q$_i$ within the waiting period (w), it is labeled as buggy and indexed (line $9$ to $11$). Alternatively, if no bug is linked to qi  during the waiting period, it is then considered clean, indexed and also stored in the clean queue (CLQ) (line $14$ to $18$). The algorithm examines the CLQ to determine if a stored change is later found to be linked to a defect. In that case,  its label is swapped to buggy and it is re-indexed with the new correct label (line $22$ to $25$).

The goal of the aforementioned \textit{online} evaluation is to obtain a comprehensive performance profile for the model, showcasing its capabilities in a genuine online learning scenario where it predicts commits based on all prior data up to each time step. However, when it comes to JITLine and JITFine evaluation, given the high computational cost associated with retraining these model, it was not feasible to retrain them at each time step. Therefore, we conducted a \textit{batched} evaluation which was recently used by Cabral \emph{et al.}~\cite{cabral2023investigation} for JIT-SDP. To carry out \textit{batched} evaluation, we randomly sampled \textit{n} time steps from all time steps, at which we retrain the \textit{batched} model. In this paper, we use $n=10$ considering our time budget, which means we update the model (or retrain in case of JITLine) at $10$ different time steps.

To evaluate \textit{batched} models, we first update the model using a training set that contains only commits with labels known by \textit{t}. After training the model, we evaluate it on the commits from $100$ subsequent time steps. As mentioned earlier, our training data is collected while keeping verification latency in account. To ensure a fair comparison between the two approaches, we re-evaluated IRJIT under this \textit{batched} evaluation methodology, allowing us to directly compare its performance with that of JITLine under similar constraints.

\subsubsection{RQ1 setup.}
The RQ1 focuses on analyzing the CPU or GPU run time of the evaluated approaches. We measure the CPU run time in seconds for all evaluations of the ORB, IRJIT, and JITLine models. For the JITFine model, we measure the GPU run time. The CPU-based experiments were conducted on a machine equipped with Intel® Core™ i7-10610U CPU @ 1.80GHz × 8 and a 16GB RAM. For JITFine, experiments were performed on a machine featuring an Intel(R) Core(TM) i7-6850K CPU @ 3.60GHz, RAM 64GB, and 2 GPUs: NVIDIA GeForce RTX 3090 (24GB) and NVIDIA GeForce RTX 4090(24GB). 

For JITLine, and JITFine we only measure the training time of the models. For IRJIT, we measure the duration of the entire experiment. This includes the time spent on indexing commits and the time taken to predict incoming commits through querying the indexed data(section~\ref{sec:MLT}). Our aim in doing so is to include the time associated with query execution and search processes in our evaluation. To estimate the run time distribution we repeat our experiment $10$ times. As we used the dataset of Cabral~\emph{et al.}~\cite{cabral2019class} we also use the waiting period suggested by them, i.e., $90$ days. 

\subsubsection{RQ2 Setup}
The RQ2 focuses on analyzing the predictive performance of IRJIT and its comparison with the evaluated baselines: ORB, JITLine and JITFine using popular online learning metrics such as G-mean, recall, and difference of recalls. Given the inherent differences in operation between online and batch processing models, our experimental setup required a methodological adjustment to facilitate a fair comparison with batch-oriented approaches like JITLine and JITFine. Therefore, IRJIT is evaluated in both \textit{online} and \textit{batched} mode for comparison with ORB and JITLine/JITFine respectively.

To build models for \textit{batched} evaluation, we segmented the data into \textit{n} batches at randomly picked time steps. We then retrain and evaluate our models on these batches, and measure performance. In our evaluations, the value of \textit{n} is set to $10$, so we train and evaluate on $10$ time steps. We ensure that IRJIT and baseline models were evaluated on the same time steps so they could be comparably evaluated against each other.

\subsubsection{RQ3 Setup}
The RQ3 focuses on evaluating the effectiveness of fine-grained line-level prediction. To accomplish this, we first acquire line-level ground-truth data using the methodology outlined in prior work~\cite{pornprasit2021jitline,jitfine,yan2020JIT}. We identify buggy lines as those altered by defect fixing commits. We used PyDriller~\cite{pydriller} to retrieve all defect-fixing commits. Subsequently, we extracted the modified or deleted lines from these commits and categorized them as buggy lines, while the remaining lines were labeled as clean.
	
Once the data was collected, we proceeded to rank the buggy lines within each predicted buggy commit from a \textit{batched} model. To rank lines, we utilized BM25 term weights or importance scores of tokens obtained from the Elasticsearch Explain API. The API generates an explanation for queries and specific documents, detailing why a particular document matches (or doesn't match) a query.
We parsed the output from the Explain API to identify tokens contributing to the prediction of a buggy commit. Subsequently, we searched for all these identified tokens within the lines added to a buggy commit and ranked them based on the occurrence frequency of distinct tokens within each line. The resulting ranked lines were then provided to developers for inspection.

\section{Experimental Results}
\subsection{RQ1. Cost-effectiveness of IRJIT}
This section discusses the findings related to RQ1, focusing on the CPU or GPU computational time of the evaluated approaches. Table~\ref{table:runtime} presents a comparison of the mean run time costs for the online models, specifically IRJIT$_{online}$ in relation to ORB, as well as the offline models including JITLine$_{batched}$ and JITFine$_{batched}$, and their comparison with IRJIT$_{batched}$.
From the Table~\ref{table:runtime}, we can observe that the highest cost is associated with DL-based approach, JITFine, followed by the ML-based approach, JITLine. ORB is the most cost-effective approach as its computational cost is less than a minute per project. This speed can be attributed to its truly online nature and its reliance solely on change metrics instead of source code.In contrast, IRJIT relies on changes to the source code, and on average, indexing these changes takes about four times as long as querying does.

\begin{table}[ht]
	\centering
	\caption{Mean run time in seconds for all evaluated approaches. JITFine experiments were run on GPU. All other experiments were executed on CPU.}
	\label{table:runtime}
	
	\begin{tabular}{l|rr|rrr}
		& \multicolumn{2}{c|}{Online} & \multicolumn{3}{c}{Batched} \\
		\hline
		Project       & 
		\small{IRJIT} &
		\small{ORB}   &
		\small{IRJIT}  &
		\small{JITLine} &
		\small{JITFine}

		\\
		\hline  
		\hline
		Brackets		  & 5444.29 & 4.97 & 	4107.71 &  {16000.47} &	{40061.84}\\
		BroadleafCommerce & 3904.26 & 3.54 &	2973.80	&  {19302.67} & {36338.51}\\
		Camel 		      & 7874.40 & 5.79 &	7043.70	&  {19645.57} &	{57354.90}\\
		Fabric8           & 801.60	& 2.38 &    1016.03 &  {23182.17} &	{49431.29}\\
		JGroups           & 611.39  & 3.74 &	607.84  &  {13544.64} &	{38360.11}\\
		Neutron           & 3279.62 & 3.56 &	2544.35 &  {17173.35} &	{37833.65}\\
		Nova              & 12443.82& 13.90 &	12221.37&  {55909.59} &	{77206.80}\\
		Npm 			  & 353.32  & 2.29 &    290.73  &  {4097.68}  &	{34963.38}\\
		Spring-integration& 659.48  & 3.07 &	779.14  &  {8736.78}  &	{34220.75}\\
		Tomcat            & 2057.91 & 4.42 &	1792.93	&  {15108.53} &	{32040.22}\\
		
		\hline
	\end{tabular}
\end{table}

\begin{figure}
	\centering
	\includegraphics[width=0.99\linewidth]{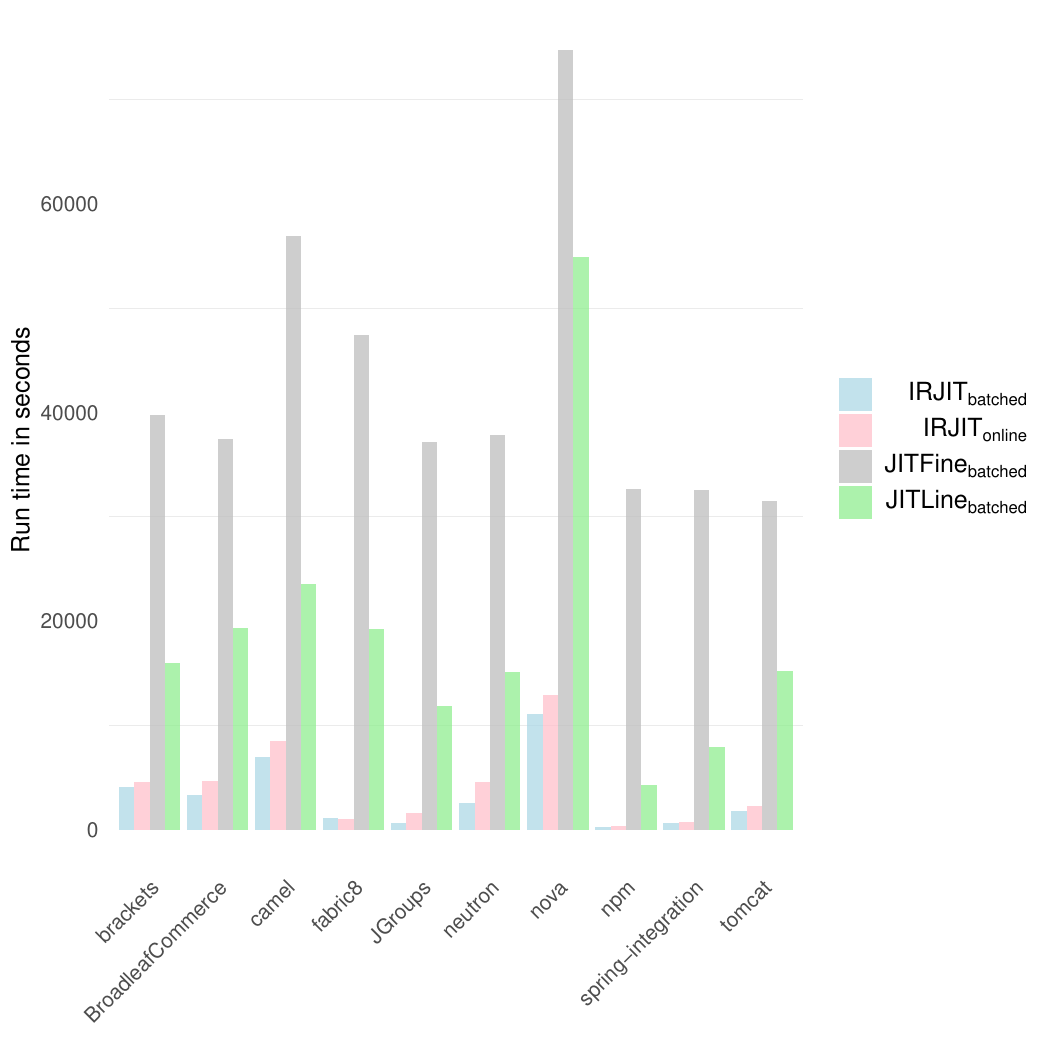}
	\caption{Comparing CPU/GPU run time of the evaluated approaches in seconds. Only JITFine was evaluated on GPU. ORB is not included in the plot because of its negligible run time.}
	\label{fig:rq3timebarplot}
\end{figure}

The computational cost of JITLine is $3$ to $23$ times higher than that of IRJIT under comparable conditions, indicating that adopting IRJIT could potentially reduce CPU run times by a factor of $3$ to $23$. The mean computational cost of JITLine consistently surpasses the maximum values of IRJIT, indicating a persistent trend of higher CPU run time requirements for JITLine across all projects. Furthermore, JITLine's scalability issues become evident as the dataset size increases. This is illustrated through Figure~\ref{fig:rq3timebarplot}, where we observe a sharp increase in the height of the bar for the Nova project. These scalability issue stems from JITLine's reliance on SMOTE to generate synthetic minority class samples to balance the class distribution. The parameters of SMOTE are tuned via the Differential Evolution (DE) algorithm, which accounts for majority of processing time. With larger datasets, the complexity of identifying the parameters and generating synthetic points increases underscoring JITLine's struggle to handle larger datasets effectively.

Similarly, the DL-based approach, JITFine, is $7$ to $112$ times more costly than IRJIT. The computation time for JITFine could be significantly higher when using a CPU. Moreover, running such an approach might not be feasible without high-end GPU hardware. For example, it is noteworthy that we were unable to execute this approach on a machine equipped with only 8GB of GPU memory; consequently, we had to upgrade to a machine that offers 24GB of GPU memory. Despite utilizing a high-end machine, the model training time for JITFine on a GPU is substantially higher than the training time for our approach on a CPU. This highlights that a cost-effective JIT-SDP approach can not only bring budgetary benefits for companies by reducing operational expenditures but might also enhance the scalability.

Our study also identified that the most resource-intensive tasks within IRJIT's workflow
is the indexing of source code changes. We observed that on average, indexing is four times more time-consuming than the combined querying and prediction process. To further illustrate this difference, we measured the time required to process a single commit, finding that indexing requires an average of $0.1$ seconds per commit, whereas querying and prediction takes only $0.04$ seconds per commit. These significant bottlenecks—indexing in IRJIT and parameter tuning in JITLine—suggest key areas for improvement for future work. Given that indexing frequency directly influences processing time, an optimized indexing strategy could potentially reduce these costs, making the system more economically viable for companies looking to implement it.

To conclude, while the computational costs of IRJIT, both in online and batched modes, are closely aligned, these costs are very high in comparison to the online approach ORB. On the other hand they are significantly lower compared to JITLine and JITFine. This suggests that researchers should focus on designing JIT-SDP approaches that operate in an online manner, eliminating the need for retraining.

\roundbox{RQ1: IRJIT is significantly more costly than ORB, which has a run time of less than $60$ seconds for each dataset. However, IRJIT reduces the CPU run time in comparison to the state-of-the-art ML and DL approaches, JITLine and JITFine, by a factor of $3$ to $112$ times. This makes IRJIT a more cost-effective option compared to the evaluated ML and DL models.}

\subsection{RQ2. Commit level predictive performance}

\begin{figure}
	\centering
	\includegraphics[width=\linewidth]{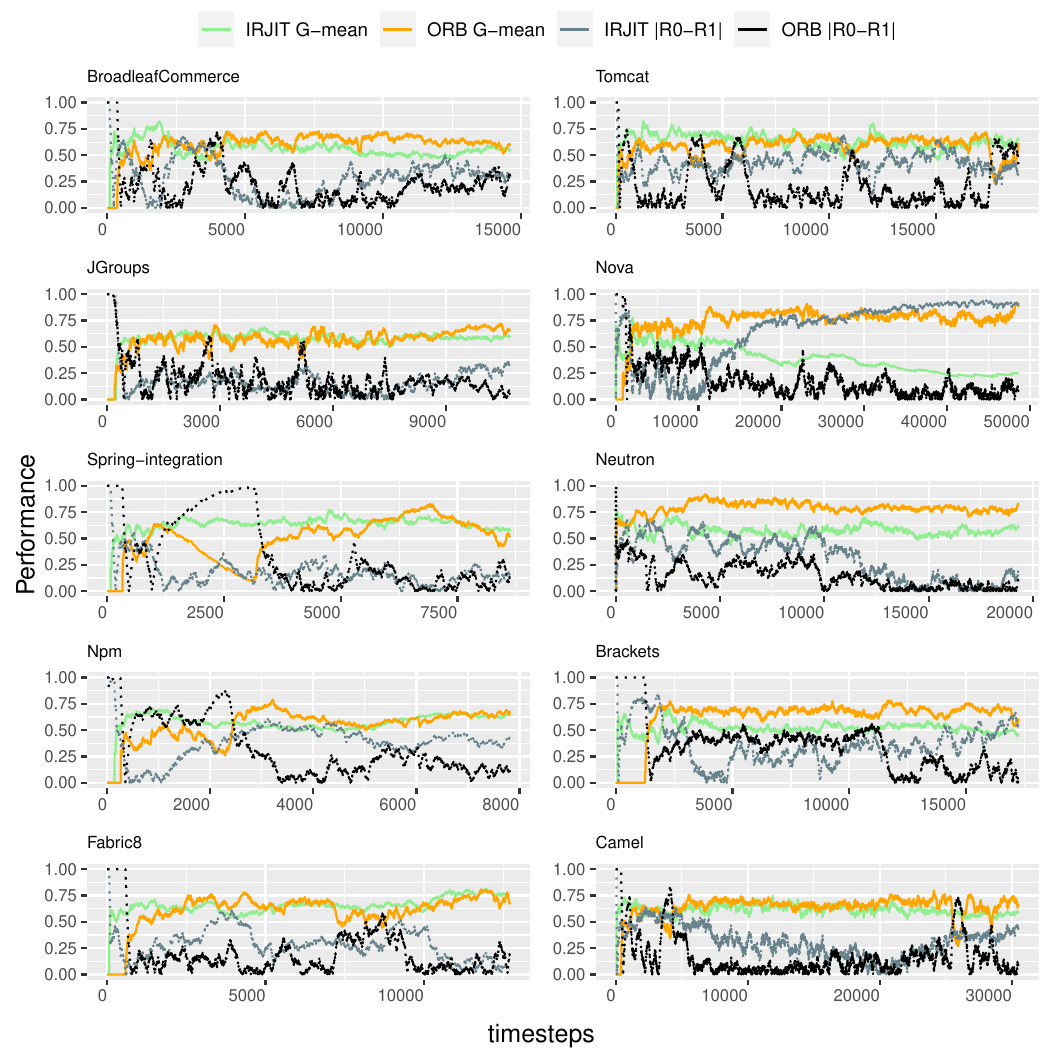}
	\caption{Comparing G-mean and difference of recalls ($|$R0-R1$|$) performance of IRJIT$_{online}$ and ORB for all datasets. X-axis shows timesteps}
	\label{fig:orbirjit}
\end{figure}

This section discusses the results of RQ2 and presents the findings of commit level predictive performance of IRJIT in online and batched evaluation. The IRJIT$_{online}$ models are compared with ORB, which is an online approach, whereas IRJIT$_{batched}$ models are compared with JITLine and JITFine.

Figure~\ref{fig:orbirjit} presents a comparison of IRJIT and ORB on the basis of G-mean and $|R_0-R_1|$ metric. The G-mean values start at $0$ but go up to $0.75$ across the projects for both approaches. Most projects achieve a G-mean of $0.5$ or higher in the very early time steps and maintain a stable G-mean, as observed in Brackets, JGroups, and Nova. On the other hand, Spring-integration and Npm experience a decline, obtaining the lowest G-mean around time step 2K. In case of Spring-integration this drop aligns with a decrease in ratio of buggy class examples. This class imbalance in turn affects the performance of ORB, although IRJIT remain unaffected. However, we do observe a constant decline in IRJIT's performance for the Nova dataset. In the remaining projects it shows a relatively steady G-mean, and surpasses ORB in $5$ out of $10$ projects including: Npm, Spring-integration, Tomcat, JGroups, and Fabric8. However, the median difference of recalls ($|R_0-R_1|$) for IRJIT is generally very high. Although, some projects such as JGroups and Spring-integration exhibit very low $|R_0-R_1|$ with median being $0.17$ and $0.19$ respectively, but it grows across datasets with the highest value ($0.60$) observed for the Nova project. In contrast, ORB achieves the lowest $|R_0-R_1|$ for BroadleafCommerce whereas the highest value $0.32$ was observed for Brackets. This suggests ORB can maintain a good recall for both buggy and clean classes across various datasets, whereas, IRJIT fails to do so in some cases.

\begin{figure}[ht]
	\centering
	
	\begin{subfigure}[b]{0.48\textwidth}
		\centering
		\includegraphics[width=\textwidth]{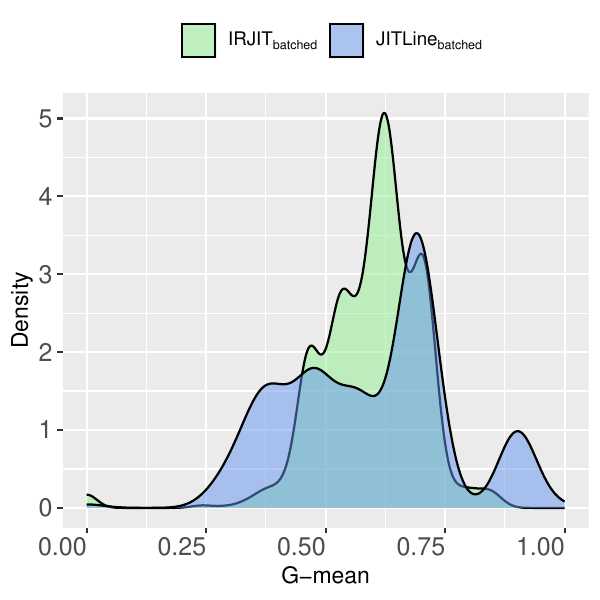}
		\caption{}
		\label{fig:irjit_jitline}
	\end{subfigure}
	\hfill 
	\begin{subfigure}[b]{0.48\textwidth}
		\centering
		\includegraphics[width=\textwidth]{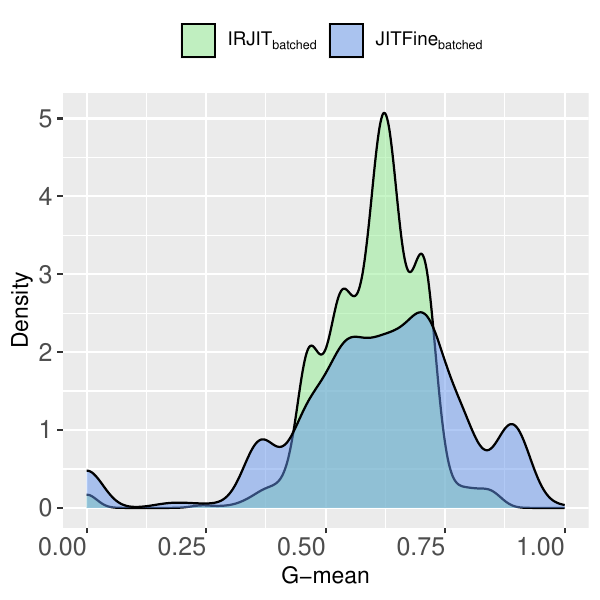}
		\caption{}
		\label{fig:irjit_jitFine}
	\end{subfigure}
	
	\caption{Density plot comparing the G-mean distribution of IRJIT$_{batched}$ with (a) JITLine$_{batched}$ and (b) JITFine$_{batched}$.}
	\label{fig:density}
\end{figure}

The overall G-mean distributions of the batched evaluation of IRJIT, JITLine and JITFine is presented in Figure~\ref{fig:density}. The density plot suggests that IRJIT's distribution leans towards a unimodal pattern, predominantly centering around a G-mean value of $0.63$. IRJIT plot is slightly skewed to the right, indicating a concentration of values around the mode with a tail extending towards the higher values. However, the tail is relatively short, suggesting a lower variability and that the values of G-mean are more consistently close to the mode.
In contrast, JITLine's distribution is bimodal, with a substantial presence of data across a broader range of G-mean values, suggesting a higher level of variability because the values of G-mean are not clustering around a central value. In case of JITFine, there are no distinct peaks in the distributions, and we observe a relatively smooth variability rather than clustered performance.
As expected, the standard deviation of IRJIT ($0.11$) is lower than both JITLine ($0.17$) and JITFine ($0.19$), confirming our observations from the plot. Nevertheless, the overlapping regions in the distributions in both Figure~\ref{fig:irjit_jitline} and ~\ref{fig:irjit_jitFine} hint at situations where both methods offer comparable results to IRJIT.

\begin{figure}
	\centering
	\includegraphics[width=\linewidth]{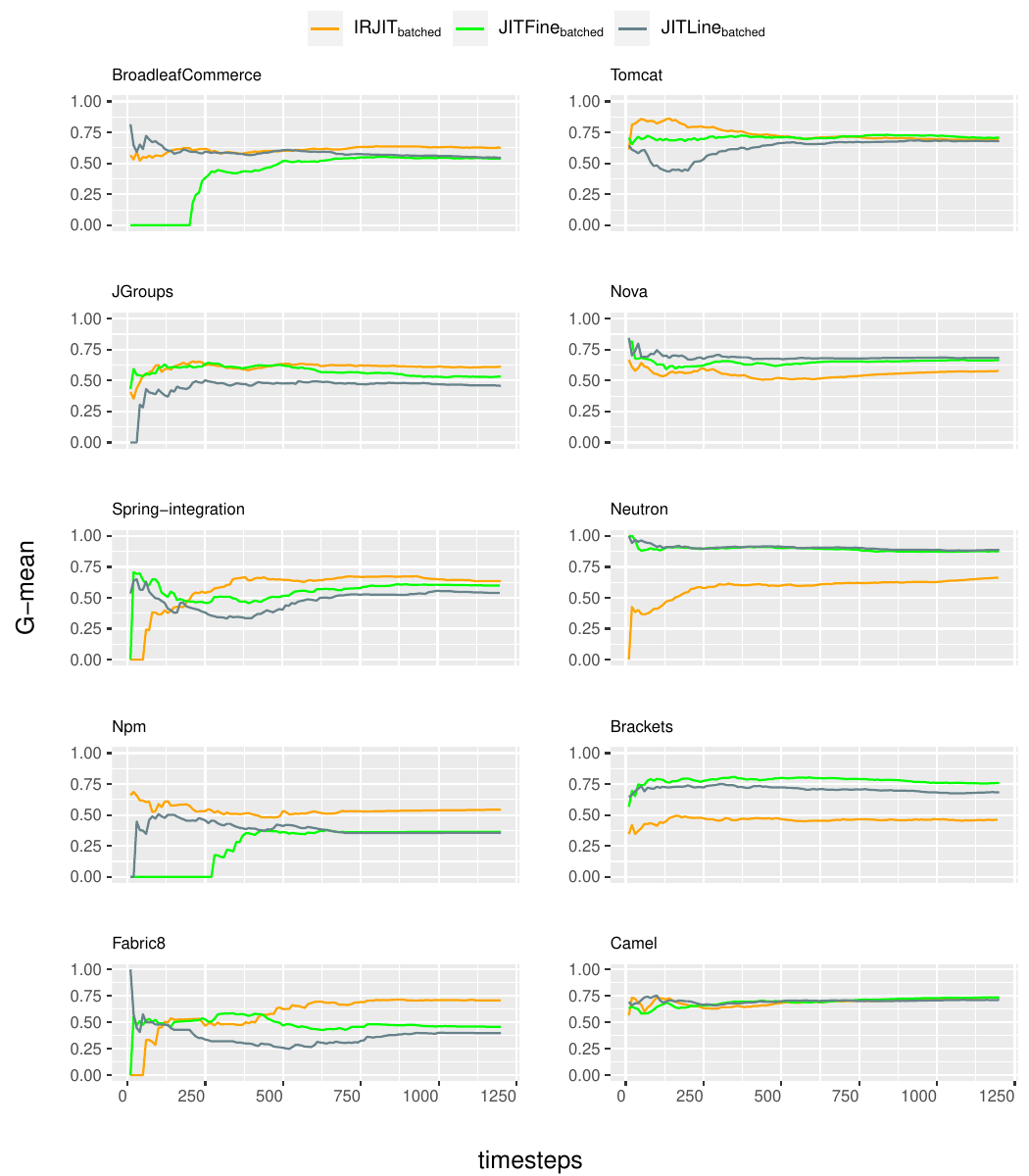}
	\caption{Commit level performance of IRJIT$_{batched}$, JITLine$_{batched}$ and JITFine$_{batched}$ models across $10$ datasets. X-axis represents time steps and Y-axis represents G-mean($\nearrow$) performance across time steps.}
	\label{fig:irjitjitlineby10}
\end{figure}

\begin{table}[ht]
	\centering
	\caption{Performance comparison of IRJIT with JITLine and JITFine using G-mean, Recall of buggy class (R$_{1}$), Recall of clean class (R$_{0}$), difference of recalls $|R_0-R_1|$, and False Alarm Rates (FAR).}
	\label{table:stats}
	
	\begin{tabular}{llccccc}
		\textbf{Approach} & \textbf{Project} & \textbf{G-mean} &
		 $\boldsymbol{R_0}$ & $\boldsymbol{R_1}$ & 
		$\boldsymbol{|R_0-R_1|}$
		 &\textbf{FAR}   \\ 
			
		\hline
		\hline
			    			& IRJIT   & 0.46 & 0.81 & 0.26 & 0.55 & 0.19 \\ 
	    			Brackets& JITLine & 0.68 & 0.92 & 0.51 & 0.41 & 0.08 \\ 
			    			& JITFine & 0.76 & 0.86 & 0.67 & 0.20 & 0.14 \\ \hline
			    			& IRJIT   & 0.54 & 0.87 & 0.34 & 0.53 & 0.13 \\ 
    		    \textbf{Npm}& JITLine & 0.36 & 0.97 & 0.13 & 0.84 & 0.03 \\ 
			    			& JITFine & 0.37 & 0.96 & 0.14 &  0.82 & 0.04  \\ \hline
			    			& IRJIT   & 0.73 & 0.70 & 0.77 & 0.07 & 0.30 \\ 
  			  \textbf{Camel}& JITLine & 0.71 & 0.91 & 0.56 & 0.35 & 0.09 \\
			    			& JITFine & 0.73 & 0.82 & 0.66 & 0.17 & 0.18 \\ \hline 
			    			& IRJIT   & 0.58 & 0.84 & 0.40 & 0.44 & 0.16 \\ 
		  		    	Nova& JITLine & 0.68 & 0.95 & 0.49 & 0.46 & 0.05 \\ 
			    			& JITFine & 0.67 & 0.90 & 0.49 & 0.41 & 0.10 \\ \hline 
			    			& IRJIT   & 0.66 & 0.60 & 0.73 & 0.14 & 0.40 \\ 
	     		  	Neutron & JITLine & 0.89 & 0.88 & 0.90 & 0.01 & 0.12 \\ 
			    			& JITFine & 0.88 & 0.86 & 0.89 & 0.03 & 0.14 \\ \hline 
			   				& IRJIT   & 0.70 & 0.54 & 0.90 & 0.37 & 0.46 \\ 
			      	  Tomcat& JITLine & 0.68 & 0.87 & 0.53 & 0.34 & 0.13 \\ 
					    	& JITFine & 0.71 & 0.79 & 0.64 & 0.15 & 0.21 \\ \hline				
			   			 	& IRJIT   & 0.63 & 0.71 & 0.56 & 0.15 & 0.29 \\
  \textbf{BroadleafCommerce}& JITLine & 0.55 & 0.94 & 0.32 & 0.62 & 0.06 \\ 
			    			& JITFine & 0.54 & 0.89 & 0.33 & 0.56 & 0.11 \\ \hline
			    			& IRJIT   & 0.71 & 0.87 & 0.57 & 0.30 & 0.13 \\ 
			\textbf{Fabric8}& JITLine & 0.40 & 0.99 & 0.16 & 0.83 & 0.01 \\ 
			    			& JITFine & 0.46 & 0.93 & 0.22 & 0.71 & 0.07 \\ \hline
			    			& IRJIT   & 0.61 & 0.65 & 0.58 & 0.07 & 0.35 \\ 
			\textbf{JGroups}& JITLine & 0.46 & 0.95 & 0.22 & 0.73 & 0.05 \\ 
							& JITFine & 0.53 & 0.82 & 0.35 & 0.47 & 0.18 \\ \hline 
							& IRJIT   & 0.63 & 0.77 & 0.52 & 0.25 & 0.23 \\ 
\textbf{Spring-integration} & JITLine & 0.54 & 0.92 & 0.32 & 0.60 & 0.08 \\ 
							& JITFine & 0.60 & 0.86 & 0.41 & 0.45 & 0.14 \\ 
		\hline
		\hline
	\end{tabular}
\end{table}

Figure~\ref{fig:irjitjitlineby10} presents a per project comparison for both approaches to reveal performance differences under specific conditions. In the figure G-mean metric is represented on the y-axis while x-axis shows time steps. Both IRJIT and baseline approaches exhibit performance peaks, but they differ in terms of magnitude and frequency. Specifically, IRJIT outperforms JITLine and JITFine on the basis of G-mean and $|R_0-R_1|$ in $6$ out of $10$ projects shown as bold in Table~\ref{table:stats}. For the Camel and BroadleafCommerce datasets, the performance trajectories of IRJIT and JITLine overlap, making it challenging to differentiate their effectiveness (see Figure~\ref{fig:irjitjitlineby10}). For example, in Camel, IRJIT undergoes steep declines in performance around time step $100$. Similarly, BroadleafCommerce shows a high G-mean in the early time steps before it drops at later time steps. For JGroups and Tomcat datasets, IRJIT is very close to JITFine.

On the other hand, JITFine and JITLine emerge as clear winners in $3$ projects showing superior G-mean performance. These projects include Nova, Neutron and Brackets, specifically IRJIT has a poor G-mean of $0.46$ for Brackets dataset. For Nova, IRJIT achieves a G-mean value of $0.58$ whereas the other two approaches reach $0.67$ and $0.68$, though the difference of recall of JITLine approach is higher than IRJIT. The highest number of false alarms are produced by IRJIT, followed by JITFine, while JITLine has the lowest FAR, signifying that it will be least wasteful of developer's effort.

To evaluate the statistical significance of the observed differences between IRJIT$_{batched}$ and the baseline models, we employed the Mann-Whitney test for independent samples. We compared the G-mean results of IRJIT$_{batched}$ and JITLine$_{batched}$ per project. We applied Bonferroni Correction to the $alpha$ values to correct for multiple comparisons. For JITLine, we were able to reject the \textit{null} hypothesis for all datasets except Camel, but for JITFine both Camel and Tomcat hypothesis could not be rejected. These findings underscore that while IRJIT shows statistically significant differences compared to ML and DL baselines, the superiority of any method is dataset-dependent, with no clear winner across the board.

\subsubsection{Impact of K}
\begin{figure}
	\centering
	\includegraphics[width=0.9\linewidth]{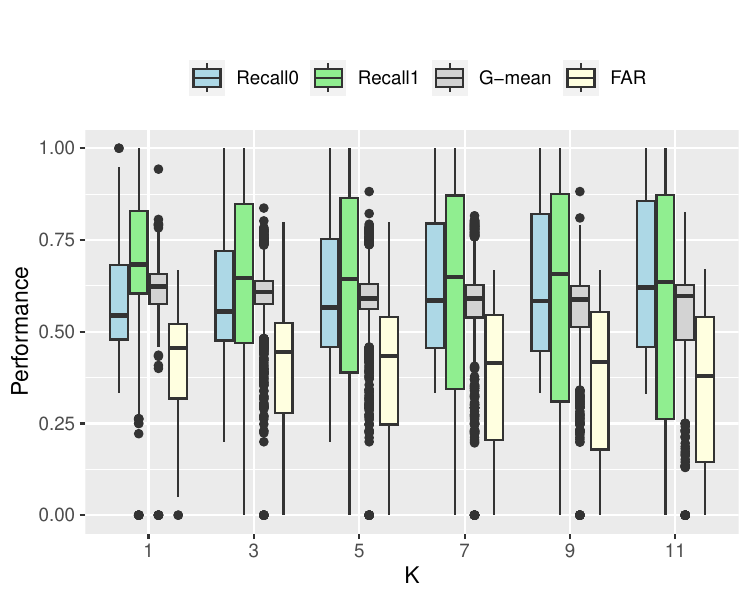}
	\caption{Performance variation of IRJIT$_{online}$ models across different values of K represented along the x-axis.}
	\label{fig:irjitkplotk}
\end{figure}

As for the \texttt{KNN} classifier we tested the model's performance for different $K$ values. Specifically, we tested for $K=1,3,5,7,9,11$ and the impact on model's performance is presented in Figure~\ref{fig:irjitkplotk}. 

In our analysis, we observe minor variations in the G-mean across different values of K. While a low $K$ leverages the precision of BM25 to accurately measure the relevance of commits, increasing the value of $K$ means that the clean class examples might dominate the voting process. This leads to an increase in $R_0$ from $0.54$ to $0.62$, as illustrated in Table~\ref{table:Knn_K}. Concurrently, we notice a decrease in FAR from $0.46$ to $0.38$ with an increase in $K$ from $1$ to $11$. While at first glance, the reduction in FAR seems favorable, it introduces significant trade-offs such as drop in $R_1$.

More concerning, however, is the effect on $|R_0-R_1|$, which jumps from $0.25$ to $0.47$. This increase suggests a growing imbalance in model performance across the two classes, highlighting the need for careful consideration in the choice of $K$. A relatively balanced recall can be achieved with smaller $K$ values at the cost of a higher FAR which we consider a reasonable choice for a JIT-SDP approach intended for early detection of buggy commits. In such contexts, prioritizing the early detection of bugs, even if it means tolerating a higher rate of false alarms, could outweigh the benefits of reducing false positives at the risk of overlooking actual bugs. These observations leave us with two choice ($K=1$ and $K=3$) for the default values for IRJIT, striking a balance between stability and performance. By choosing $K=3$ as the default $K$ value, we obtain a prediction based on the unanimous agreement between the top $3$ matches, which is presumably more reliable than a single match.

\begin{table}[ht]
	\centering
	\caption{Median performance across different values of K using Recall${_0}$ (R${_0}$), Recall${_1}$ (R${_1}$), $|R_0-R_1|$, G-mean and False Alarm Rate (FAR)}
	\label{table:Knn_K}
	
	\begin{tabular}{lrrrrrr}
		Metric/K &  {K=1} &  {K=3}&  {K=5} & {K=7} &   {K=9} &  {K=11} \\\hline \hline
		
		\textbf{Recall${_0}$}     & 0.54 & 0.56    &  0.57 & 0.58 & 0.58 & 0.62 \\
		\textbf{Recall${_1}$}     & 0.68 & 0.65    &	0.64 & 0.65 & 0.66  & 0.64 \\
		$\boldsymbol{|R_0-R_1|}$  & 0.25 & 0.34    &  0.40 & 0.45 & 0.46 & 0.47 \\
		\textbf{G-mean}           & 0.62 & 0.61    &	0.59 & 0.59 & 0.59  & 0.60 \\
		\textbf{FAR}              & 0.46 & 0.44    &	0.43 & 0.42 & 0.42  & 0.38 \\
		
		\hline
	\end{tabular}
\end{table}

In conclusion, while no definitive statistical distinction exists between the two methodologies, their performance varies depending on the dataset. While JITLine occasionally excels, IRJIT often matches or surpasses JITLine in certain projects. This variability emphasizes the importance of understanding dataset characteristics, suggesting that relying solely on one method might not always be the best strategy. Further research could shed light on the specific conditions favoring each approach.

\roundbox{RQ2: IRJIT shows a competitive predictive performance at commit-level, achieving a higher G-mean than ORB in $5$ out of $10$ projects. It also outperforms the ML and DL baselines JITLine and JITFine in $6$ out of $10$ projects, considering both G-mean and difference of recalls ($|R_0-R_1|$) of buggy and clean class. In only $3$ out of $10$ projects the  baseline approaches performed better than IRJIT.}

\subsection{RQ3. Line level performance}
This section presents the results of RQ3, which evaluates line-level prediction performance of IRJIT in comparison to baseline fine-grained approaches, including JITLine, JITFine, and the N-Gram model by Yan et al. In Section~\ref{sec:ranklines}, we previously detailed the utilization of term weights from the Elasticsearch Explain API~\cite{explainapi} to rank buggy lines. After ranking lines we compute four line-level evaluation metrics described in section~\ref{sec:metrics}. The remainder of this section presents the results of our line-level evaluation.

\begin{figure}
	\centering
	\includegraphics[width=0.99\linewidth]{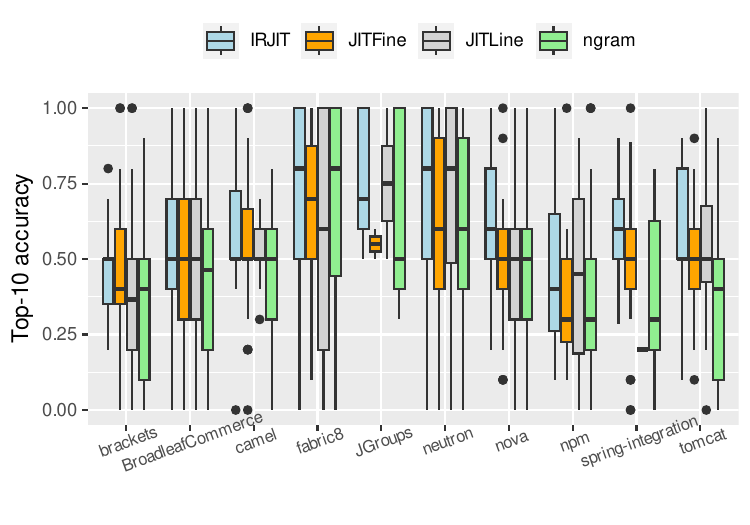}
	\caption{Comparing line level performance of IRJIT,  JITFine, JITLine, and n-gram across $10$ datasets Top-K accuracy($\nearrow$)}
	\label{fig:topk}
\end{figure}

All four line-level metrics except IFA range from $0$ to $1$. To remain consistent with prior work~\cite{pornprasit2021jitline,yan2020JIT}, we use the top $10$ lines to compute Top-K accuracy. The median Top-10 accuracy of IRJIT and JITLine is $0.64$ and $0.61$ respectively. A high ratio suggests that many of the buggy lines are ranked at the top. As shown in Figure~\ref{fig:topk}, IRJIT has a higher median top-k accuracy than JITFine, JITLine, and N-gram in $6$, $4$, and $7$ projects respectively. The relatively smaller interquartile range for IRJIT also suggests that it exhibits a more consistent top-k performance across projects. JITLine exhibits more variability in its accuracy, especially for fabric8, npm and neutron. The overall variability for JITLine measured using standard deviation is $0.30$, which is higher than that of IRJIT ($0.26$). All approaches have outliers, but they are more pronounced in the case of JITFine, suggesting that while JITFine might work well for specific datasets, it might not be consistently  effective for all datasets.

\begin{figure}
	\centering
	\includegraphics[width=0.99\linewidth]{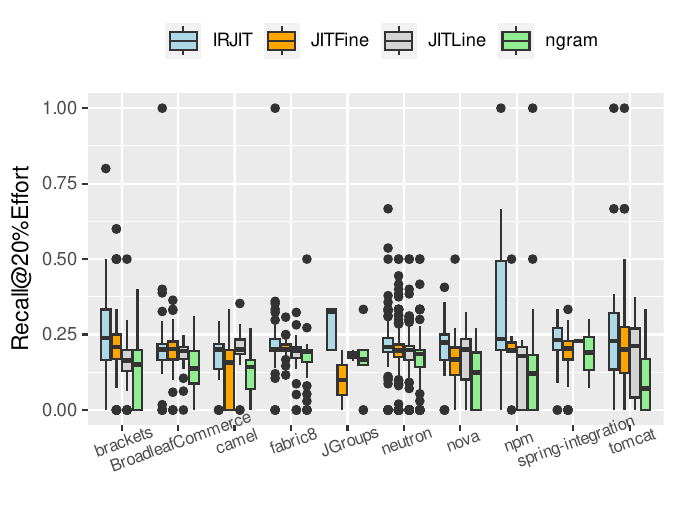}
	\caption{Comparing line level performance of IRJIT, JITFine, JITLine, and N-gram across $10$ datasets using Recall@20\%Effort($\nearrow$) }
	\label{fig:recall}
\end{figure}

\begin{figure}
	\centering
	\includegraphics[width=0.99\linewidth]{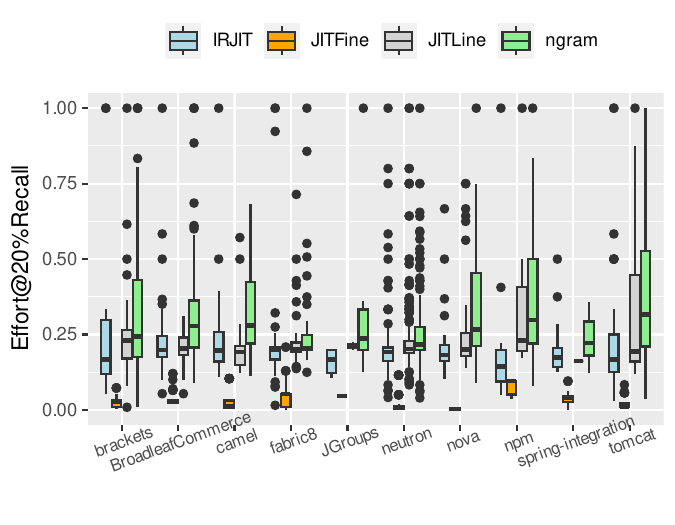}
	\caption{Comparing line level performance of IRJIT, JITFine, JITLine, and n-gram across $10$ datasets using Effort@20\%Recall($\searrow$)}
	\label{fig:effort}
\end{figure}

A significant portion of defects in a software application can often be traced back to a relatively small portion of the codebase. We use Recall@20\%Effort to measure the ratio of actual buggy lines that can be found given only top 20\% changed lines of code. A high value of Recall@20\%Effort indicates that
an approach can rank many actual buggy lines at the top
and many actual buggy lines can be found given a fixed
amount of effort. On the other hand, a low value of this metric indicates that many clean lines are in the top 20\%
LOC and developers need to spend more effort to identify
defective lines. Figure~\ref{fig:recall} shows that the  Recall@20\%Effort for all approaches is under $0.25$ with the median for IRJIT being slightly higher than the other approaches in most datasets. This suggests IRJIT can rank more buggy lines among the top 20\% of lines.

\textit{Effort@20\%Recall} measures the amount of effort made by the developers to identify top20\% of buggy lines in a commit. A higher value suggests that more effort is needed so a lower value is better.
Figure~\ref{fig:effort} presents Effort@20\%Recall for each project.
JITFine has the lowest values of Effort@20\%Recall signifying that out of all approaches it requires the least amount of effort in order to find the same amount of actual defective lines. On the other hand, N-gram has the highest effort values so developers using this approach will spend more effort to find the actual 20\% defective lines of a defective commit. Although IRJIT does not beat JITFine, but it requires 5\% and 18\% less effort than the baseline approaches, JITLine and N-gram, respectively. JITLine's effort varies more drastically from one project to another, and in some projects the outliers are much more prominent, such as neutron and fabric, but it is close to IRJIT with median Effort@20\%Recall values of $0.20$ and $0.19$ respectively.

\begin{figure}
	\centering
	\includegraphics[width=0.99\linewidth]{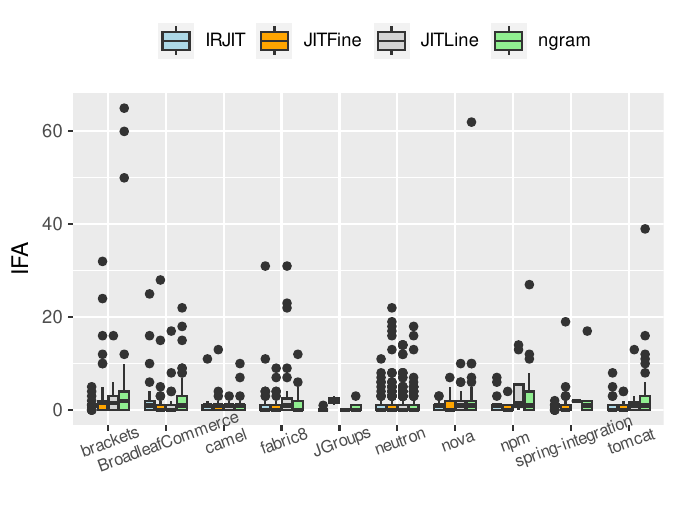}
	\caption{Comparing line level performance of IRJIT,  JITFine, JITLine, and n-gram across $10$ datasets using IFA($\searrow$)}
	\label{fig:ifa}
\end{figure}

\textit{Initial False Alarm (IFA)} measures the amount of clean lines of a commit that need to be inspected before identifying an actual buggy line. A low IFA value indicates that few clean lines are ranked at the top, while a high IFA value indicates that developers will spend unnecessary effort on clean lines.
The median IFA for all approaches is $0$ indicating that a buggy line is always ranked as the top location.
However, as shown in Figure~\ref{fig:ifa}, IRJIT has a slightly lower IFA for some projects. It indicates that buggy lines are always ranked at the top so developers need to inspect fewer clean lines until finding the first actual defective line for a given commit. JITLine not only has a higher IFA but also more pronounced outliers, making it particularly well-suited for some contexts but less effective in others. To confirm if the observed differences between any of the metrics were statistically significant or not we conducted Mann-Whitney test. For JITLine, we found that the differences between top-k and IFA are not statistically significant ($p-value = 0.241$ and $p-value = 0.996$), but those between Recall@20\%Effort and Effort@20\%Recall ($p-value = 1.265e-10$ and $p-value = 2.49e-11$) are significant. 
For JITFine the differences for Top-K and Effort@20\%Recall between IRJIT and JITFine are statistically significant ($p-value < 2.2e-16$).

\roundbox{RQ3: IRJIT is better than JITLine, JITFine and the N-gram approach considering Top-K metrics. In terms of Effort@20\%Recall, JITFine outperformed IRJIT and other approaches. The Mann-Whitney test confirms that the observed differences are statistically significant. In terms of IFA all approaches are similar.}

\section{Discussion} 
Our study indicates that while sophisticated machine learning models might be more effective in certain contexts, their efficiency in real-time or rapidly changing environments is still in question. Given that not all organizations or researchers have access to the computational power or other resources required to train complex models, simple approaches are the way forward. Our study shows that IRJIT can balance between performance and cost in such scenarios, democratizing access to JIT models for a broader audience.
A key advantage of IRJIT is that it offers explainability at two distinct levels. At the commit-level, it provides context of the predictions by directly linking to the contributing documents i.e., commits that contributed to the prediction. At line level, it ranks buggy lines of an identified buggy commit, guiding developers to dedicate their limited effort in certain parts of the code for identifying bugs.

The IRJIT model is online and was trained only on within project data. The model initially predicts clean until it sees buggy examples. Tabassum\emph{et al.} showed that cross project data was helpful in the initial phase of the project when there was no or little within project training data available. Future studies may include cross project data to improve model accuracy.
We subjected IRJIT to rigorous testing on extensive codebases, including prominent projects, such as nova and camel, to validate IRJIT's practicality and effectiveness when applied to projects with substantial codebases. Furthermore, it is worth noting that our experiments were conducted on a machine with modest computing resources. This underscores the versatility and practical applicability of IRJIT in various environments, demonstrating that it can deliver valuable results even without high-end hardware.

\subsection{Implications}
\textit{Researchers should focus on designing simple approaches for JIT-SDP.} Our findings reveal that in terms of performance, IRJIT closely rivals a state-of-the-art machine learning approach. However, what sets IRJIT apart is its inherent simplicity compared to JITLine. IRJIT relies solely on source code changes which are readily available in the project's source code repository. It uses basic IR techniques that can be replicated with ease, even by individuals with limited expertise, thereby enhancing the accessibility of IRJIT. Our study demonstrates that IR-based methods are competitive out of the box, suggesting that further tuning and using them in combination with other project-specific features (e.g., metrics) can bridge the performance gaps.

\textit{Researchers should evaluate JIT approaches taking verification latency into account.}
Based on RQ1, our study further emphasizes that, when deciding which JIT-SDP approach to adopt, it is important to investigate them taking veriﬁcation latency into account. Prior studies often relied on the traditional 80:20 split for the JIT evaluation. Following recent work~\cite{cabral2019class,tabassum2020investigation}, we conduct an evaluation at different points in time while taking verification latency into account. Through this evaluation, we emphasize two crucial points: 1) defect prediction approaches exhibit varying performance over time, and 2) some approaches may be impractical for resource-constrained real-world development settings. Therefore, we advocate for research that aligns with the practical needs and constraints faced by developers in the field. Future defect prediction studies should adopt evaluation methodologies that accurately reflect the real-world application of JIT prediction from a development perspective.

\textit{Researchers must ensure that their JIT approach is suitable for online deployment and does not require considerable retraining.}
In general, JIT approaches with lower computational cost are better suited for seamless online deployment because they do not slow down the agile development. Our work demonstrates the practicality of IRJIT, which operates online without necessitating model tuning or retraining. Instead, it incrementally updates indexes as new commits arrive, making it highly adaptive and scalable. In contrast, the baseline approach, JITLine, relies on resource-intensive retraining involving techniques like SMOTE and DE. Consequently, JITLine is computationally more expensive for online deployment and potentially slows down the overall prediction process.
However, different projects and requirements may warrant different approaches. It is worth noting that the choice between IRJIT and JITLine may vary depending on specific project requirements and available resources. While IRJIT prioritizes efficiency and adaptability, JITLine might offer superior performance on certain datasets. Developers with ample resources and a strong emphasis on model accuracy may opt for JITLine over IRJIT.

\subsection{Limitations and Opportunities for Future Work}

\subsubsection{Reliance on SZZ} We rely on SZZ algorithm to obtain the ground truth dataset for evaluation. We as part of the software engineering community are aware of the shortcomings of SZZ algorithm that it only identifies large defects that get reported to the issue tracking system and therefore are important enough to be fixed. Other kinds of datasets such as ManySStuBs4J~\cite{manysstubs} are not widely available so we rely on a dataset extracted with the help of SZZ. In the area of defect prediction, researchers commonly rely on SZZ e.g., it was used to extract the datasets in recent JIT studies ~\cite{ISSTA2021,pornprasit2021jitline,cabral2019class,tabassum2020investigation}. We therefore think that it is out of the scope of our paper to design a new way of identifying buggy commits. Instead, we implore the software engineering community to come together and take up the challenge of improving the accuracy of SZZ algorithm or propose new ways of identifying representative buggy commits. 

\subsubsection{Usability of Explanations}
JITLine employs SMOTE to invent documents, making it more difficult to trace predictions back to the original documents involved in commit-level prediction. When IRJIT flags a commit, it provides a concrete, context-relevant example as an explanation. Although we have not yet conducted a user study to assess the effectiveness of IRJIT explanations, we aim to explore this aspect in future work through a user study.

For line-level predictions, JITLine trains a local LIME model~\cite{ribeiro2016should}, which adds additional cost on top of the original model. Similar to JITFine, IRJIT improves upon JITLine by eliminating the need for two separate models for prediction and localization. It uses a unified IR model to identify a buggy commit and rank its lines by defectiveness. This integration ensures that explanations are generated promptly alongside predictions, essential for real-time application in development workflows. Our decision prioritizes simplicity and speed, making IRJIT more suitable in practice.

\subsection{Impact of Class Imbalance and Concept Drift}
Class imbalance and concept drift are well-known problems in SDP literature. However, we did not investigate their impact on IRJIT. We believe that our model's continuous update mechanism at each timestep captures the latest trends and patterns in the data, and addresses the aforementioned issues to a significant extent. In other words, since the training set is continuously updated, it is more likely to have similar characteristics as the test set for constructing accurate models.
However, we recognize the importance of a thorough examination of these factors and are open to exploring their impact on IRJIT's performance in future work. We believe that experimenting with class imbalance and concept drift mitigation techniques such as online SMOTE could potentially benefit IRJIT by enhancing its performance.
\section{Threats to Validity}
Here we discuss threats that impact the credibility and generalization of our findings. The choice of datasets used for evaluation should be representative of typical software projects, or sampling bias is introduced, affecting the \textit{internal validity} of results. However, we counteract the sampling bias by using an existing dataset, which includes ten diverse projects encompassing three popular programming languages and multiple domains. Moreover, our datasets' different sizes and complexities also ensure that our findings are not restricted to certain types of projects, hence improving the \textit{external validity}.  

We also acknowledge that comparisons against additional baselines such as DeepJIT and CC2Vec could add more value to our work. To this end, we only compared IRJIT with four prior approaches. However, it is important to note that the chosen baselines, JITLine and JITFine, have been evaluated against these models in prior studies and demonstrated superior performance. This was a key factor in our decision to use them as baselines for our research.

The choice of features or information retrieval methods may influence the effectiveness of IRJIT. Currently, IRJIT collects and uses \texttt{lines added} by default because the probability of introducing a bug is higher with the addition of new lines, as suggested by Purushothaman \emph{et al.}~\cite{test}, and hence spotting bugs might be easier with only \texttt{lines added} field. Variations in hyper-parameter settings of the IRJIT model can impact results. Robust hyperparameter tuning is necessary to minimize this threat. We tested IRJIT on different values of K in the range $1$ to $11$.

The \textit{conclusion validity} of a study is affected by the choice of performance metrics because that affects how we measure the success of an approach. Following, several prior works, we decided to use G-mean and difference of recalls because it is the most widely accepted metric in online learning. Using multiple metrics or considering alternative measures can provide a more comprehensive assessment. Therefore, at the line level, we use JITLine's proposed metrics. Furthermore, we optimized JITLine for G-mean instead of AUC to keep the comparison between the two approaches consistent. Our study also assumes a verification latency waiting period of $90$ days which means a commit becomes a training example after $90$ days from commit time have passed. However, projects that are not frequently modified may require longer waiting period to obtain confidence on the label of committed software changes.

\section{Conclusion}
This study investigated JIT-SDP in a realistic online learning scenario using $10$ open-source datasets. We showed that our approach, IRJIT, has a comparable performance to the state-of-the-art approach ORB but a higher run time cost. We also showed that state-of-the-art ML and DL approaches JITLine and JITFine, are cost-prohibitive in an online JIT-SDP setting, due to their expensive training process. To address this problem, we propose IRJIT, an information retrieval approach for defect prediction that relies on incremental model updates. Our approach saves the run time cost by a factor of $3$ to $112$ times against the sophisticated baselines. It also achieves a competitive G-mean performance
at the commit-level that surpasses JITLine and JITFine, in $6$ out of $10$ projects. Notably, IRJIT achieves JITLine level performance in an online setting without using SMOTE. IRJIT is explainable, as it makes predictions by linking current changes to historical buggy changes, allowing developers to benefit from the rich context inherent in the evolution of the codebase. It also ranks the changed lines by riskiness, making it easier for developers to identify bugs with minimal effort. The line-level evaluation shows that IRJIT has a higher Top-k performance than the aforementioned approaches and an N-gram approach, whereas JITFine was best in terms of Effort@20\%Recall. Our study implies that if practitioners want to save time and obtain predictions in real-time, then IRJIT is a good option as it provides competitive performance with much lower run time compared to ML and DL baselines. Moreover, as IRJIT operates directly on the source code without needing additional hand-crafted features, it saves computational resources and time, leading to faster predictions. In a future study we will incorporate additional open-source projects and will compare different online JIT-SDP approaches with IRJIT.

\section*{Data Availability Statement}
The dataset used in evaluation, the source code to evaluate approaches and the R scripts to plot the graphs for this study are available at: \\https://github.com/Hareem-E-Sahar/eseval\_online

\section*{Conflict of Interest Statement}
The authors of this article declared that they have no conflict of interest.


\bibliographystyle{spmpsci}
\bibliography{sample-base}


\end{document}